\documentclass[aip,graphicx,preprint,amsmath,amssymb]{revtex4-1}
\usepackage[utf8]{inputenc}
\usepackage[T1]{fontenc}
\usepackage{mathptmx}
\usepackage{etoolbox}
\usepackage{braket}
\usepackage[version=3]{mhchem} %
\usepackage[caption=false]{subfig}
\usepackage{graphicx} %
\usepackage{ragged2e}
\usepackage{xcolor}
\usepackage{xfrac}

\usepackage[colorlinks, allcolors=blue]{hyperref}
\usepackage{multirow}
\newcommand{\refeq}[1]{{Eq.~(\ref{#1})}}
\newcommand{\reffig}[1]{{Fig.~\ref{#1}}}
\newcommand{\refsec}[1]{{Sec.~\ref{#1}}}
\newcommand{\reftab}[1]{{Table~\ref{#1}}}

\draft %
\makeatletter
\def\@email#1#2{%
 \endgroup
 \patchcmd{\titleblock@produce}
  {\frontmatter@RRAPformat}
  {\frontmatter@RRAPformat{\produce@RRAP{*#1\href{mailto:#1}{#2}}}\frontmatter@RRAPformat}
  {}{}
}%
\makeatother
\begin{document}

\title{A Single Twist-Angle Selection Method for the Electronic Structure of Bilayer Materials} %

\author{Ryan A. Baker}
    \email{bakerry9@msu.edu}
\affiliation{Department of Chemistry, Michigan State University, East Lansing, MI}
\author{William Z. Van Benschoten}
\affiliation{Department of Chemistry, Michigan State University, East Lansing, MI}
\author{James J. Shepherd}
    \email{sheph158@msu.edu}
\affiliation{Department of Chemistry, Michigan State University, East Lansing, MI}

\date{\today}

\begin{abstract}
Structure factor twist averaging (sfTA) is a newer method that has been shown to reproduce twist-averaged (TA) CCSD energies for bulk systems at a low computational cost.
In this work, we extend this method for the treatment of low-dimensional materials in the form of two variants: paired sfTA and binding sfTA.
These variants affect which twist angles are used in the sfTA protocol, as well as how the special twist angle is selected, namely by using the binding structure factor.
These changes are meant to incorporate the binding interaction into the twist-angle selection algorithm within sfTA.
Both variants are tested on a variety of bilayer systems, and the resulting binding correlation energies are compared to original sfTA results.
We show that the variants are able to produce results approaching TA, with binding sfTA producing the most accurate energies.
We also use contour plots of the test systems to show that these improvements are most likely caused by a cancellation of errors. 
\end{abstract}

\maketitle %

\section{Introduction}

    The development of many-body methods for solids, and general electron correlation methods in particular, is a research area focused primarily on performing accurate calculations at a low computational cost.\cite{shi_many-body_2023}
    A major challenge for the development of these methods lies in dealing with finite size errors (FSEs).
    When finite supercells are used to model an infinite solid, long-range effects cannot be treated explicitly due to the limiting size of the supercell, leading to FSEs in the energy.\cite{drummond_finite-size_2008}
    One method of treating FSEs is to obtain energies for large supercells run on a dense $k$-mesh, then extrapolate the energy to the thermodynamic limit, i.e. the infinite supercell limit.\cite{gruber_ab_2018,liao_communication_2016,marsman_second-order_2009,booth_towards_2013,mihm_power_2021}
    However, good extrapolations require some minimum sized supercell to capture enough long-range interactions.
    If the desired method has a large computational cost, such as the $\mathcal{O}\left(N^6\right)$ scaling in coupled cluster singles and doubles (CCSD), then achieving the supercell size needed for a good extrapolation may not be practical or possible due to limitations of memory and/or compute time.\cite{zhang_coupled_2019}
    For this reason, techniques that can provide accurate correlation energies for many types of materials by mitigating FSE at small supercells are highly sought after.
    
    One method of mitigating FSE that has been shown to be effective for bulk materials is twist averaging (TA).\cite{drummond_finite-size_2008,lin_twist-averaged_2001}
    The TA protocol averages a set of calculations at different offsets to the $\Gamma$-centered $k$-mesh, known as twist angles, and uses the average as a representation of a denser k-mesh rather than directly increasing the mesh density.
    While effective, TA still requires at least one calculation at the desired level of theory for each twist angle, leading to large computational costs that scale linearly with the number of twist angles, $N_s$.
    A recent method called structure factor twist averaging (sfTA) can achieve TA-level results for the correlation energy with only a single high-level calculation.\cite{mihm_shortcut_2021}
    For a single system, sfTA performs calculations over a set of random twist angles, similar to TA, but using a computationally inexpensive method compared to the target accuracy.
    For example, in this work, we use 2nd-order M{\o}ller-Plesset perturbation theory (MP2), which when using natural orbitals has a cost scaling of only $\mathcal{O}\left(N^4\right)$, two orders of magnitude less than CCSD.\cite{gruneis_natural_2011}
    Next, the electronic transition structure factor, a property of the correlation energy that relates the correlation energy to $t$-amplitudes in the wavefunction, is used to select a special twist angle.\cite{liao_communication_2016}
    The transition structure factor at this special twist angle should be a good approximation of the twist-averaged transition structure factor, and, therefore, a good approximation for the TA correlation energy.
    The sfTA special twist angle is then used for a single high-level calculation with the end result being the sfTA energy, which has been shown to closely approximate the TA CCSD result for bulk metals, semiconductors, and insulators.\cite{mihm_shortcut_2021}

    While sfTA has been shown to be very effective for bulk materials, it has yet to be tested on low-dimensional materials.
    Low-dimensional materials are materials that exhibit periodic behavior in fewer than three dimensions, such as sheets (2D) or chains (1D).
    These materials often exhibit different and unique electronic properties compared to their analogous bulk counterparts, so investigating their electronic structure is of interest to materials scientists.\cite{novoselov_2d_2016}
    In particular, some of the most interesting chemistry and physics associated with these materials comes from interactions with other materials, such as small molecules binding to surfaces in catalysis, or the interactions of layered materials that we study in this work.
    In the original paper, the sfTA method was only used to obtain total energies for bulk materials, and the authors did not explore what changes may be necessary to adapt it for low-dimensional materials, let alone for potential interaction energies.\cite{mihm_shortcut_2021}
    Adapting \textit{ab initio} methods and techniques developed for bulk materials to low-dimensional materials and their interactions with other systems is a common approach, but often requires special considerations. \cite{yadav_first-principles_2023}
    For example, many modern exchange-correlation functionals for density functional theory struggle to accurately capture the van der Waals interactions that lead to binding in layered materials, so alternate variants of these functionals are needed instead.\cite{klimes_perspective_2012, silvestrelli_van_2008}
    While van der Waals interaction naturally emerge from wavefunction based methods like coupled cluster and quantum Monte Carlo, these methods still require other types of adaptations to accurately treat 2D materials. \cite{mostaani_quantum_2015}
    With these considerations, we seek to adapt sfTA to obtain interaction energies of low-dimensional materials, test the accuracy of these adaptations, and explore why these adaptations work.

    In this paper, we use the original sfTA method to obtain the interlayer binding correlation energy for a series of bilayer systems to see if it maintains the same accuracy to TA as seen for bulk systems.\cite{mihm_shortcut_2021}
    We also test two sfTA variants that have been modified to account for the binding interaction in some way, to see if the adaptations results in more accurate binding correlation energies.
    One of the variants, binding sfTA, uses the binding transition structure factor, which is a difference of the structure factors between the mono- (1L) and bilayer (2L) systems used for calculating the binding energy.
    The original sfTA protocol is then employed, using the binding structure factor in lieu of the transition structure factor.
    A consequence of this approach is that both the 1L and 2L systems need to have the same set of twist angles, or else calculating the binding structure factor will be impossible.
    The second variant, paired sfTA, simply uses the twist angles from binding sfTA, and performs sfTA independently on 1L and 2L.
    This approach allows for the 1L and 2L systems to select the same twist angle.
    This was previously impossible because for original sfTA, new sets of random twist angles were generated for each system.
    We show that all variants, including the original sfTA, result in a similar accuracy compared to TA for the lone 1L and 2L systems as was seen for bulk materials.
    We also show that original sfTA shows similar accuracy for the binding correlation energy, while paired and binding sfTA lead to noticeable improvements, with binding sfTA being the most accurate overall.
    Finally, we use contour maps of the binding energy as a function of the two-dimensional twist angle offset to explore why the set of paired twist angles used for paired and binding sfTA lead to improvements in the binding correlation energy.
    Using this approach, we see a clear cancellation of errors, meaning any twist angle, so long as it is held constant between 1L and 2L, should result in a binding correlation energy close to the TA energy.
    
\section{Methods}
    \label{sec:methods}
    \subsection{Electron Correlation}
        \label{sec:methods:corr}
        The correlation energy ($E_{\mathrm{corr}}$) is the difference between the total electronic energy ($E_{\mathrm{Total}}$) and the energy from the Hartree-Fock (HF) approximation ($E_{\mathrm{HF}}$):
        \begin{equation}
            E_{\mathrm{corr.}}=E_{\mathrm{total}}-E_{\mathrm{HF}}
        \end{equation}
        In the HF approximation, explicit electron-electron potential interactions are neglected and approximated by a one-electron mean field potential instead.\cite{fock_naherungsmethode_1930}
        Correlation energy methods seek to recover the electronic interactions neglected by the HF approximation.
        They often differ from one another in precisely how much correlation they include.\cite{stanton_why_1997}
        In general, methods that account for more electronic correlation are both more accurate and more computationally expensive and are referred to here as "high-level", while less accurate methods require less computational power, and are referred to here as "low-level".
        Some methods try to transfer information from low-level calculations to high-level calculations, like in this work.
        
        In this work, we use 2nd-order M{\o}ller-Plesset Perturbation theory (MP2) and coupled cluster singles and doubles (CCSD) as low- and high-level correlation energy methods respectively.
        Using the notation from Liao and Gr{\"u}neis, $E_{\mathrm{corr}}$ is found as:
        \begin{equation}
            E_{\mathrm{corr}}=\sum_{ij}\sum_{ab}T_{ij}^{ab}v_{ij}^{ab}
        \end{equation}
        where $ij$ and $ab$ are occupied and unoccupied orbital indices respectively; $v_{ij}^{ab}$ is the two-electron electron repulsion integral; and the wavefunction $t$-amplitudes are given by
        $T_{ij}^{ab}$.\cite{liao_communication_2016} 
        
        For CCSD, the $t$-amplitudes are derived from the cluster operator of the exponential ansatz
        \begin{equation}
            \ket{\psi}=e^{\hat{T}}\ket{\phi}
            \label{eq:exponential_ansatz}
        \end{equation}
        where $\psi$ and $\phi$ are the CCSD and HF wavefunctions, respectively, and $\hat{T}$ is the cluster operator.
        The cluster operator in CCSD includes single and double excitations, as well as nonlinear combinations of single excitations represented by the $t$-amplitudes: $T_{ij}^{ab}=t_{ij}^{ab}+t_i^at_j^b$.
        The nonlinear equations used for CCSD need to be solved iteratively, which adds a considerable computational cost.

        To contrast, the $t$-amplitudes for MP2 come from a single equation:
        \begin{equation}
            T_{ij}^{ab}=t_{ij}^{ab}=\frac{v_{ij}^{ab}}{(\epsilon_i+\epsilon_j)-(\epsilon_a+\epsilon_b)}
            \label{eq:tijab_MP2}
        \end{equation}
        where the nonlinear $t_i^at_j^b$ term vanishes due to Brillouin's theorem, $\epsilon$ are HF eigenvalues, and $v_{ij}^{ab}$ are electron repulsion integrals.\cite{attila_szabo_modern_1989}
        The fact that this is only a single non-iterative equation leads to a significantly reduced cost of $\mathcal{O}(N^5)$, which can be further reduced to $\mathcal{O}(N^4)$ with the use of natural orbitals, compared to the $\mathcal{O}(N^6)$ scaling from CCSD.\cite{zhang_coupled_2019, gruneis_natural_2011}
        However, the change in accuracy is quite noticeable.
        For example, perturbative methods like MP2 can lead to unphysical energy divergences in metals.\cite{shepherd_many-body_2013}
        The inaccurate behavior seen in MP2 is acceptable here primarily because it is used only as an intermediate in the sfTA method to get a CCSD energy.
        Due to a similar focus on two-electron interactions, as well as the practical consideration that MP2 is the first step in a CCSD calculation for the software used in this work, these two methods seem a natural pair for comparison.
        It should be noted that in practice, other computationally inexpensive methods like the random phase approximation (RPA) could be used in place of MP2.\cite{mihm_shortcut_2021}
        
    \subsection{The Electronic Transition Structure Factor}
        \label{sec:methods:tsf}
        The expression for the transition structure factor using terms from coupled cluster has been detailed by Liao and Gr{\"u}neis. \cite{liao_communication_2016}
        Using the wavefunction $t$-amplitudes from the calculations of the correlation methods, the transition structure factor can be obtained by:
        \begin{equation}
            S(\mathbf{G})=\sum_{ij}\sum_{ab}T_{ij}^{ab}( 2C_i^a(\mathbf{G})C_b^{j*}(\mathbf{G})-C_i^b(\mathbf{G})C_a^{j*}(\mathbf{G}))
            \label{eq:sg_liao_gruneis}
        \end{equation}
        where $S(\mathbf{G})$ is the transition structure factor, $\mathbf{G}$ is the momentum transer vector (while $G$ is its magnitude), and $C_i^a(\mathbf{G})$ is a coefficient from the Fourier transformed co-densities of the occupied orbital $\phi_i$ and the virtual orbital $\phi_a$ for the given $\mathbf{G}$.
        In practice, the electron repulsion integrals are also represented using these same intermediate terms.
        
        The transition structure factor  is related to the correlation energy by the relationship:\cite{liao_communication_2016}
        \begin{equation}
            E_{\mathrm{corr}}=\sum^{\prime}_{\mathbf{G}} v(\mathbf{G})S(\mathbf{G}).
            \label{eq:ecorr_structure_factor}
        \end{equation}
        In this formalism, $E_{\mathrm{corr}}$ is the correlation energy; $v(\mathbf{G})$ is the reciprocal space Coulomb integral; and the sum is over all momentum transfer vectors $\mathbf{G}$, with the prime denoting that $\mathbf{G}$ outside our energy cutoffs are naturally excluded.

        The Coulomb integral $v(\mathbf{G})$ in this work is evaluated using the mean potential method:\cite{schafer_sampling_2024}
        \begin{equation}
            v(\mathbf{G})\approx \frac{1}{N_{\mathrm{part.}}}\sum_{n=1}^{N_{\mathrm{part.}}}v_n(\mathbf{G})=\frac{1}{N_{\mathrm{part.}}}\sum_{n=1}^{N_{\mathrm{part.}}}\frac{1}{\Omega_{\mathrm{BZ}}^n}\int_{\Omega_{\mathrm{BZ}}^n}\mathrm{d}^3\mathbf{q}\frac{4\pi}{(\mathbf{G}+\mathbf{q})^2}.
            \label{eq:mean_pot}
        \end{equation}
        For this method, the volume of the Brillouin zone ($\Omega_{\mathrm{BZ}}$) is partitioned into $N_{\mathrm{part.}}$ subspaces ($\Omega_{\mathrm{BZ}}^{n}$), determined by the dimensions of the cell, which are integrated over independently from one another.
        In practice, $N_{\mathrm{part.}}$ can become quite large with little impact to computational cost.
        This treatment of the Coulomb potential reduces quadrature errors of the Coulomb singularity for anisotropic systems, like those used in this work.\cite{schafer_sampling_2024}
        
    \subsection{Interlayer Binding Energies}
        \label{sec:methods:e_bind}
        The interlayer binding energy for a bilayer system is given by the thermodynamic relationship
        \begin{equation}
            E_{\mathrm{Binding}}=E_{\mathrm{2L}}-2E_{\mathrm{1L}}
            \label{eq:binding_energy}
        \end{equation}
        where 1L and 2L denote the mono- and bilayer.
        So in practice, to find a single binding energy, one must calculate energies for both 1L and 2L independently, then use \refeq{eq:binding_energy} to get the binding energy.
        In this paper, we focus on correlation energy methods, so all interlayer binding energies are reported as binding correlation energies.

    \subsection{Twist Averaging}
        \label{sec:methods:twist_averaging}
        The twist-averaging (TA) protocol applies a series of $N_s$ random twist-angle offsets ($k_s$) to the $\Gamma$-centered $k$-mesh.
        These twist angles translate the entire $k$-mesh by a relatively small amount within the unit cell in a random direction in reciprocal space.
        From here, all $N_s$ twist angles are run with the desired method, and the TA energy is calculated as the average of the resulting energies:
        \begin{equation}
            \braket{E}_{k_s}=\frac{1}{N_s}\sum_{k_s}^{N_s}E_{k_s}
            \label{eq:TA_energy}
        \end{equation}
        where $E_{k_s}$ is the energy at a single $k_s$ and $\braket{E}_{k_s}$ is the TA energy averaged across all $k_s$.
        Twist averaging has been shown to reduce finite-size effects for bulk materials and is used in conjunction with extrapolation, but since all twists are at the desired level of theory, this can be quite computationally expensive.\cite{lin_twist-averaged_2001}
        
        In this work, we use 100 twist angles for $N_s$, as this is consistent with previous work.\cite{mihm_shortcut_2021}
        Here, we use a two-dimensional twist angle in the plane of the material.
        These twists are generated within a range of 1 fractional unit, covering the entire symmetry of a single cell.
        For example, most systems have twists generated between [-0.5, 0.5), but some, like C (graphene), used [0.0, 1.0).
        Regardless, these ranges are equivalent due to symmetry.
        Additionally, we use a quasi-random Sobol' series to obtain our twists, as this allows for a more regular distribution across $k$-space while maintaining some characteristic random variable properties.\cite{sobol_distribution_1967}
        The quasi-random sequence was chosen over a more traditional pseudo-random sequence to reduce the chance of points becoming clustered near an extreme value, which would skew the TA result, and by extension, affect the interpretation of the accuracy of all methods involved.

    \subsection{Structure Factor Twist Averaging}
        \label{sec:methods:sfTA}
        Structure factor twist averaging (sfTA) seeks to reproduce the TA result at a reduced computational cost.
        The first step is to perform 100 low-level (MP2) calculations.
        Next, the twist-averaged transition structure factor is calculated, very similar to the twist-averaged energy from \refeq{eq:TA_energy}:
        \begin{equation}
            \braket{S(\mathbf{G})}_{k_s}=\frac{1}{N_s}\sum_{k_s}^{N_s}S_{k_s}(\mathbf{G}).
            \label{eq:TA_SF}
        \end{equation}
        where $\braket{S(\mathbf{G})}_{k_s}$ is the average transition structure factor across all twist angles and $S_{k_s}(\mathbf{G})$ is the transition structure factor at the twist angle $k_s$.
        Then, for each twist, the residual to the twist averaged transition structure factor is calculated:
        \begin{equation}
            r_{k_s}=\sum_{G}\left|\braket{S(G)}_{k_s}-S_{k_s}(G)\right|^2.
            \label{eq:SF_residual}
        \end{equation}
        where $r_{k_s}$ is the residual, and the sum runs over unique values of $G$.
        The residual is a measure of how close a particular structure factor is to the twist-averaged structure factor.
        The twist with the smallest residual difference, known as the special twist angle, is then used to run a single high-level calculation (CCSD) to get the sfTA energy.
        For TA, we would need to run all $N_s$ twist angles at the high-level method.
        This unnecessary for sfTA, and since we use 100 twists for $N_s$, there is a \textasciitilde100-fold reduction of cost compared to TA.
        This method is repeated for both 1L and 2L, and compared to the TA binding energy calculated using the relationship in \refeq{eq:binding_energy}.

        When using sfTA for the interlayer binding energy, it is important to note that the set of twist angles used is generated independently for each system, meaning that for a corresponding pair of 1L and 2L systems, the odds that they have even a single twist angle in common are so low they can be safely disregarded.
        This means that sfTA cannot pick the same twist angle for 1L and 2L, and that we also cannot calculate binding energies for individual twist angles.
        These limitations are addressed in different ways for the variants developed and tested in this work: paired and binding sfTA.

    \subsection{Paired sfTA}
        \label{sec:methods:paired_sfta}
        For paired sfTA, we generate only a single set of random twist angles, which is then used for both systems.
        In other words, each twist angle now has a pair of calculations corresponding to 1L and 2L, which can be used to calculate the binding energy at each twist.
        Because of this, we can calculate the TA binding energy (and its associated error) using \refeq{eq:TA_energy}, rather than propagating.
        
        From here, the procedure is largely identical to original sfTA as described in \refsec{sec:methods:sfTA}.
        Both 1L and 2L have their special twist angles calculated independently from one another, and the final energies for each system are used with \refeq{eq:binding_energy} to get the paired sfTA energy.
        Notably, this modification means it is now possible, though not guaranteed, that the same twist angle can be independently selected for both 1L and 2L, which was effectively impossible before.
    \subsection{Binding sfTA}
        \label{sec:methods:binding_sfta}

        \begin{figure}
            \centering
            \includegraphics[width=\linewidth]{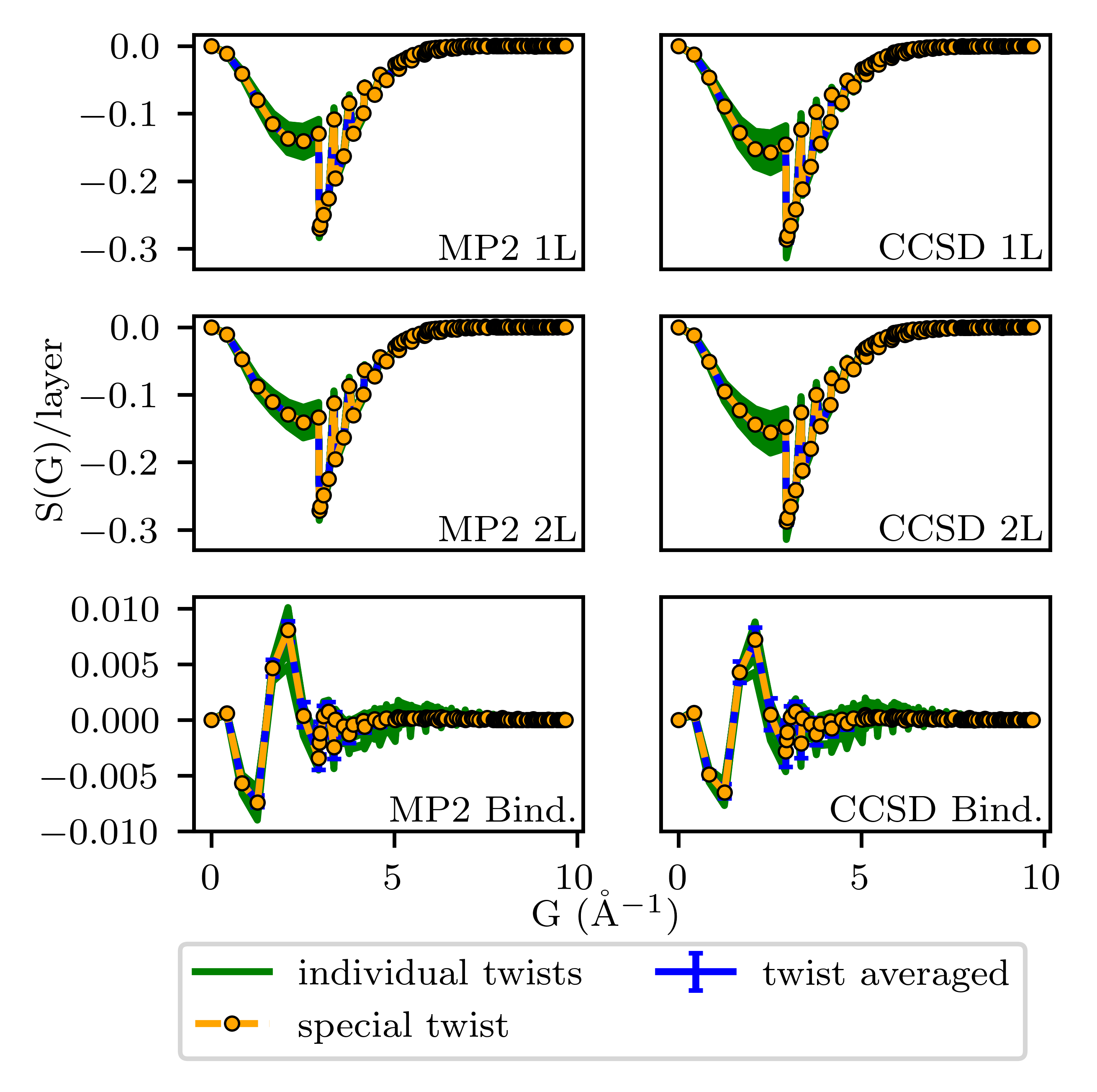}
            \caption{
                Example construction of the binding transition structure factor $S(G)$ with respect to the magnitude of the momentum transfer vector $G$ from mono- (1L) and bilayer (2L) structure factors for $1\times1$ graphene bilayer.
                Notice the shape is very consistent between MP2 and CCSD, and the special twist angle is very close to twist-averaged in all cases.
                }
            \label{fig:sg_ex}
        \end{figure}

        Binding sfTA uses the same twist angle offsets for the 1L and 2L systems, just like for paired sfTA, but also modifies the twist angle selection process.
        The process is modified by defining a new binding transition structure $S_{\mathrm{Bind.}}(\mathbf{G})$ in the same manner as the binding energy in \refeq{eq:binding_energy}:
        \begin{equation}
            S_{\mathrm{Bind.}}(\mathbf{G})=S_{\mathrm{2L}}(\mathbf{G})-2S_{\mathrm{1L}}(\mathbf{G}).
            \label{eq:binding_SG}
        \end{equation}
        The construction of the binding transition structure factor from the 1L and 2L structure factors is shown in \reffig{fig:sg_ex}.
        From here, both the 1L and 2L subsystems are run using the low-level method (MP2), and the residual differences are calculated by simply substituting the transition structure factor in \refeq{eq:SF_residual} for the binding transition structure factor.
        The binding twist angle selection process is visualized in \reffig{fig:sg_ex}.
        Once the binding special twist angle is found, both the 1L and 2L systems are rerun using the high-level method.
        One consequence of this is that the special twist angle will be the same for both 1L and 2L, which was not guaranteed for the previous paired sfTA method.

    \subsection{Numerical Analysis}
        \label{sec:methods:num_analysis}
        Figures were plotted using \texttt{matplotlib 3.9.1} with \texttt{Python 3.11.5}. 
        Box and whisker plots were made using \texttt{matplotlib.pyplot.boxplot}.        
        For box and whisker plots, the boxes encompass the middle 50\% of data from the first quartile (Q1) to the third quartile (Q3).
        This range from Q1 to Q3 is the interquartile range  (IQR).
        The whiskers extend past Q1 or Q3 to 1.5 times the IQR in either direction, and any points past that are deemed outliers, marked as gray circles.
        Contour plots were made using \texttt{matplotlib.pyplot.tricontour}.
        To increase resolution of contour plots, twist angle data was mapped into other regions of the rhombus using four-fold reciprocal lattice vectors.
        Plots along the high-symmetry $\Gamma\rightarrow M\rightarrow K$ path accompany the contour plots to show detail and are constructed from $2^7+1\ (129)$ additional CCSD calculations which are not a part of any other set in this work. 
        
        We also report the mean absolute error (MAE) throughout to measure the difference of the sfTA variants to the TA reference results:
        \begin{equation}
            \mathrm{MAE} = \frac{1}{n}\sum_{i=1}^{n}\left|E_{\mathrm{sfTA}}^{\left(i\right)}-E_{\mathrm{TA}}^{\left(i\right)}\right|
            \label{eq:MAE}
        \end{equation}
        where $E_{\mathrm{TA}}^{\left(i\right)}$ and $E_{\mathrm{sfTA}}^{\left(i\right)}$ are the reported TA and sfTA energies, respectively, for the $i$-th system, and $n$ is the number of systems within the relevant test set.
        The MAE is an aggregate measure of how closely our results match the TA results, assisting with qualitative analysis.
\section{Calculation Details}
    \label{sec:calc_details}
    \subsection{System Details}
        \label{sec:calc_details:sys_details}
        Materials were chosen to sample a range of potential electronic structures, while having clear bilayer structure and a feasible valence electron count.
        Materials were grouped into a "small test set" (\refsec{sec:results:small}) if they had a low system size and nonmetallic atoms, while all other systems were grouped into a larger "challenging test set" (\refsec{sec:results:challenge}).
        Optimized bilayer structures came from a few sources.
        \ce{C} and \ce{BN} structures were provided by collaborators (see Acknowledgments).
        \ce{Si} and \ce{SiC} were constructed from data provided by literature sources.\cite{padilha_free-standing_2015,willander_silicon_2006}
        \ce{GaN} was obtained from the materials project.\cite{jain_commentary_2013} 
        The following systems were obtained using the NOMAD online structure database.\cite{scheidgen_nomad_2023}
        \ce{MoS2} came from the Alexandria database.\cite{wang_symmetry-based_2023}
        \ce{MoSe2}, \ce{Bi}, \ce{Pb}, \ce{Al}, and \ce{Sn} all came from the open quantum materials database (OQMD).\cite{kirklin_open_2015}
        
        All systems belong to the $P6_3/mmc$ space group, the same hexagonal group as graphene.
        Systems are planar when possible, with the exceptions of \ce{MoS2} and \ce{MoSe2} due to their inherent out-of-plane structure from the dichalcogenide links, and \ce{SiC} which is buckled.
        Additionally, most systems have AB stacking with the exceptions that \ce{Si} has AA stacking, and the stacking of \ce{SiC} is derived from the 6-H polymorph.
        The structures described for \ce{C}, \ce{BN}, \ce{SiC}, \ce{MoS2}, and \ce{MoSe2} are representative of physically realized materials, while structures for \ce{Si}, \ce{GaN}, \ce{Bi}, \ce{Pb}, \ce{Al}, and \ce{Sn} may not be physical, as most experimental descriptions of these materials' mono- or multilayers involve epitaxial growth.
        We have chosen to include these structures in their current state anyway to reduce the number of factors that might influence the results, allowing us to focus on our methods.
        
        Structures were initially obtained and/or constructed as $1\times1$ bilayers.
        Interlayer distances for bilayers were kept at the same values found in the optimized structures.
        For all systems, the vacuum in the $z$-dimension was extended to 15 \AA.
        The monolayer structures used in this work were formed by removing one of the layers in the optimized bilayer structure.
        Three systems (\ce{C}, \ce{Si}, and \ce{BN}) were also run using a $2\times2$ supercell.
        Lattice constants and interlayer distances are available in the supplementary information.

    \subsection{Software}
        \label{sec:calc_details:software}
        All calculations were performed using the Vienna \textit{ab initio} simulation package (\texttt{VASP 6.3}) coupled with coupled cluster for solids (\texttt{Cc4S}).\cite{kresse_ab_1993,kresse_efficiency_1996,kresse_efficient_1996,liao_communication_2016,gruber_ab_2018}
        The projector augmented-wave method (PAW) was used with a plane wave basis set.\cite{Blochl_projector_1994}
        The HF orbitals were diagonalized within \texttt{VASP}, and MP2 natural orbitals were used for all MP2 and CCSD calculations.\cite{gruneis_natural_2011}
        Structural manipulation was performed primarily using \texttt{VESTA} (visualization for electronic and structural analysis).\cite{momma_vesta_2011}
    \subsection{Input Parameters}
        \label{sec:calc_details:input}
    Calculations required an explicit plane wave energy cutoff (ENCUT), auxiliary plane wave basis-set cutoff (ENCUTGW), and number of bands to include in the basis (NBANDS).
    All calculations use a Perdew-Burke-Ernzerhof (PBE) pseudopotential optimized using the GW approximation.\cite{perdew_generalized_1996,hedin_new_1965}
    ENCUT values were set at approximately 1.3 times the largest recommended value (ENMAX) among atoms in the pseudopotential files provided in the \texttt{VASP} distribution.
    ENCUTGW values were set to two-thirds this value.
    NBANDS were automatically set to 8 bands per occupied electron pair, which, considering most of the $1\times1$ systems have 8 electrons, leads to most systems having NBANDS of 32.
    All systems were run using a $1\times1$ $k$-mesh supercell, and some systems were also run as $2\times2$, where noted.
    
    \subsection{Twist Angle Generation}
        \label{sec:calc_detals:twist_angles}
    Twists angles were generated using a quasi-random number generator in \texttt{Python} as described in \refsec{sec:methods:twist_averaging}.
    Twist angles are the same between $1\times1$ and $2\times2$ structures of \ce{C}, \ce{Si}, and \ce{BN}.
    For each material, the twist angles used for the 2L subsystem of that material are the same across all sfTA variants.
    Twist angles for 1L subsystem were regenerated using a new seed for original sfTA, but are the same for paired and binding sfTA.
    For Bi and Al, 6 of the twists (for each system) were not able to converge in the initial DFT step in less than 10 000 steps.
    This initial failure would cascade, affecting the final energies and leading to clear outliers.
    To correct this, those twists were replaced with other twists obtained by extending the quasi-random sequence using the same seed, to minimize the impact to twist angle selection.

\section{Results}
     \label{sec:results}
     \subsection{Comparing sfTA Variants for Small Test Systems}
        \label{sec:results:small}
        \begin{figure*}
            \centering
            \includegraphics[width=\linewidth]{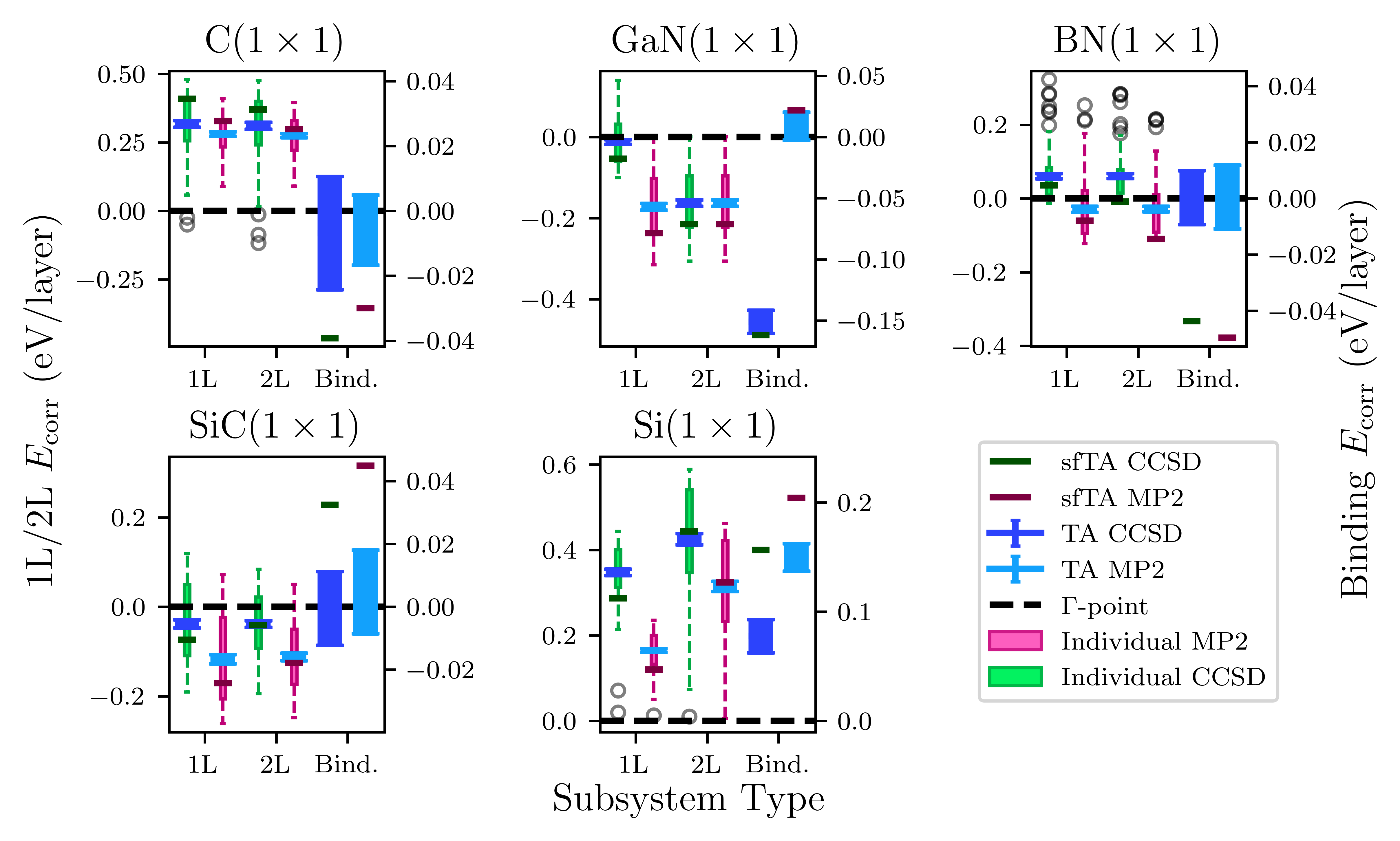}
            \caption{
                Monolayer (1L), bilayer (2L), and binding (Bind.) correlation energies for comparison of original sfTA to TA, as shown for five systems: \ce{C}, \ce{GaN}, \ce{BN}, \ce{SiC}, and \ce{Si}. 
                All energies are shown as a difference to the $\Gamma$-point energy (black dashed line).
                Each pair of box-and-whisker plots corresponds to the individual twist results for CCSD (green) and MP2 (pink).
                TA results (with standard error) are shown as dark blue (CCSD) and light blue (MP2) horizontal boxes.
                TA energies and errors for the binding energy are propagated from 1L and 2L.
                The sfTA results are shown as dark green (CCSD) and dark pink (MP2) lines.
                Binding energies are plotted on the right-hand axis for clarity.
                This figure shows original sfTA is close to TA in some cases, but is generally not in agreement within one standard error.
                }
            \label{fig:set1_original_sfta}
        \end{figure*}
        
        \begin{figure*}
            \centering
            \includegraphics[width=\linewidth]{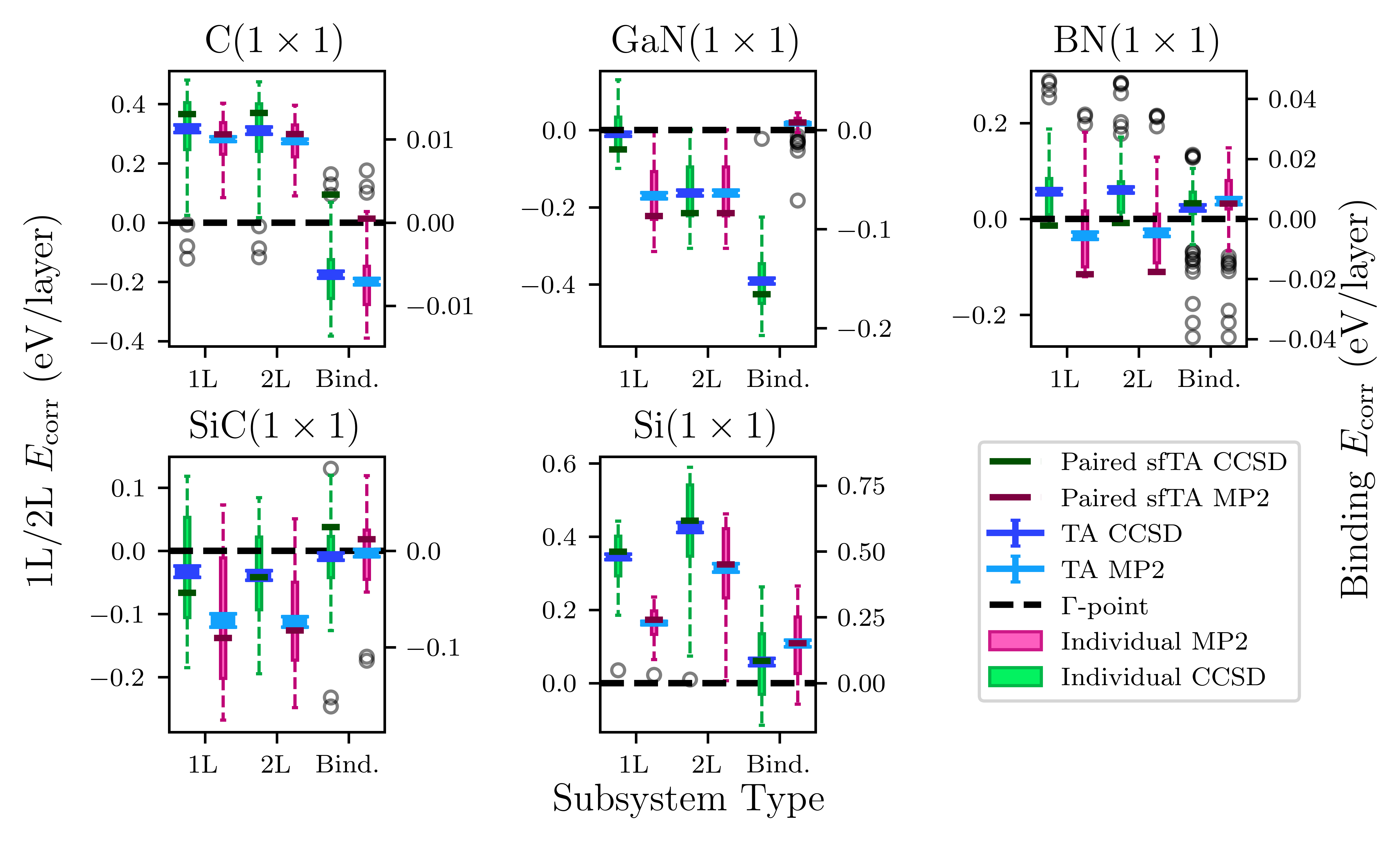}
            \caption{
                Monolayer (1L), bilayer (2L), and binding (Bind.) correlation energies for comparison of paired sfTA to TA, as shown for the five test systems: \ce{C}, \ce{GaN}, \ce{BN}, \ce{SiC}, and \ce{Si}. 
                All energies are shown as the difference to the $\Gamma$-point energy (black dashed line).
                Each pair of box-and-whisker plots corresponds to the individual twist results for CCSD (green) and MP2 (pink).
                TA results (with standard error) are shown as dark blue (CCSD) and light blue (MP2) horizontal boxes.
                The sfTA results are shown as dark green (CCSD) and plum violet (MP2).
                Binding energies are plotted on the right-hand axis for clarity.
                This figure shows that paired sfTA leads to closer agreement to TA than what was seen for original sfTA (\reffig{fig:set1_original_sfta}).
                }
            \label{fig:set1_paired_sfta}
        \end{figure*}
        
        \begin{figure*}
            \centering
            \includegraphics[width=\linewidth]{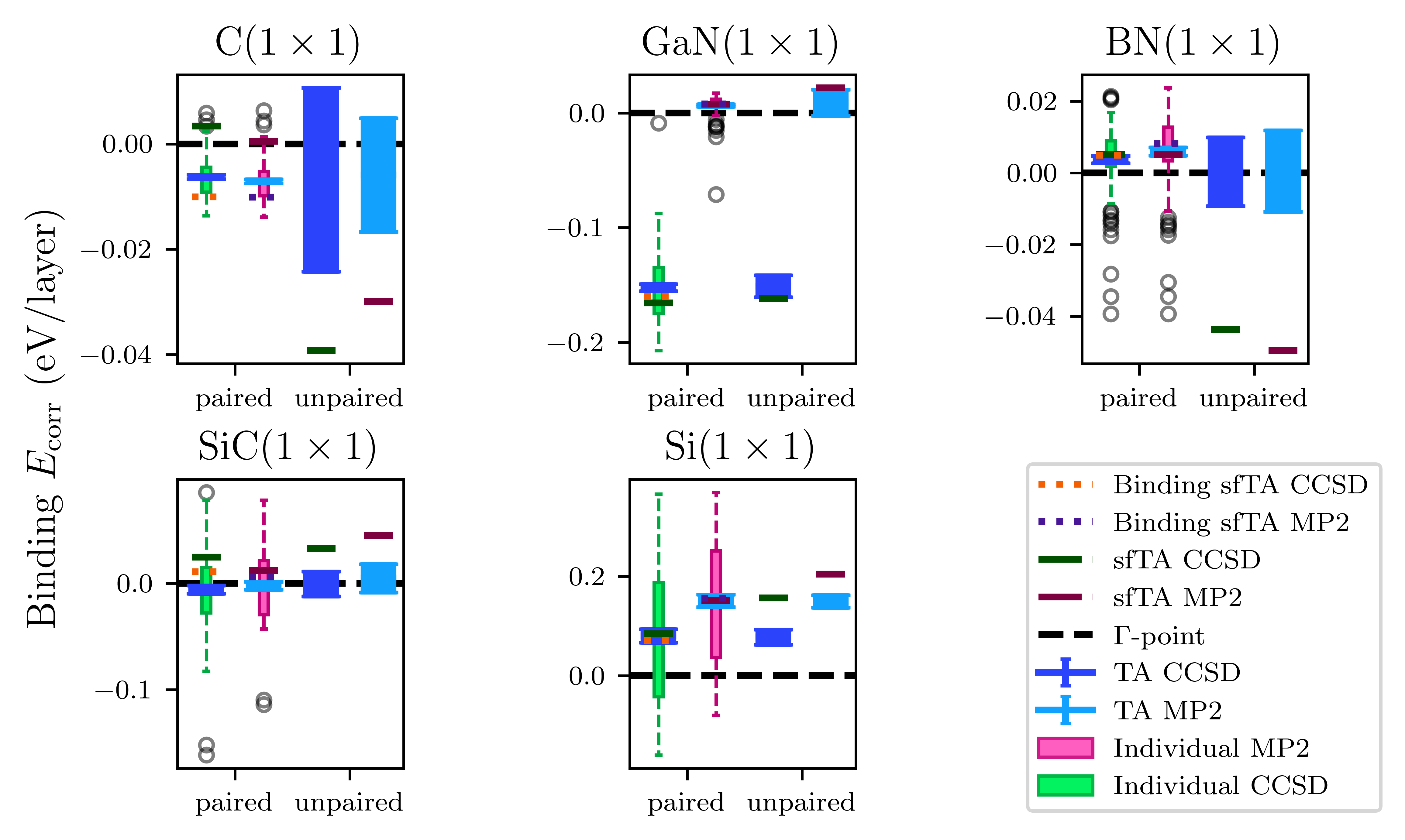}
            \caption{
                Binding correlation energy for comparison across both sfTA variants for the five test systems: \ce{C}, \ce{GaN}, \ce{BN}, \ce{SiC}, and \ce{Si}.
                All energies are shown as a difference to the $\Gamma$-point energy (black dashed line).
                Paired and binding sfTA data are grouped as paired methods (left), since they use the same twist angles throughout, while original sfTA data is marked as the unpaired method (right).
                The pair of box-and-whisker plots corresponds to the individual twist results for CCSD (green) and MP2 (pink).
                TA results (with standard error) for paired methods are shown as dark blue (CCSD) and light blue (MP2) horizontal boxes.
                TA values and errors for original sfTA are propagated from 1L and 2L.
                Original sfTA and paired sfTA are marked as dark green (CCSD) and plum violet (MP2) lines.
                Binding sfTA is shown as dotted orange (CCSD) and violet (MP2) lines.
                This figure shows that for the binding correlation energy, paired sfTA is generally closer to the TA result, while binding sfTA is closer still.
            }
            \label{fig:set1_binding_plus_all}
        \end{figure*}

        \begin{table}[]
            \caption{
                Mean absolute errors (MAE) with uncertainty in the last decimal place for 1L, 2L, and binding correlation energies for all test sets, in units of meV/atom. 
                MAE for 10 bulk test systems from original sfTA paper was 3.9(8) meV/atom.\cite{mihm_shortcut_2021} 
            }
            \label{tab:MAE}
            \begin{tabular}{c|ccc|ccc|ccc}
                \hline
                Variant  & \multicolumn{3}{c|}{sfTA} & \multicolumn{3}{c|}{Paired sfTA} & \multicolumn{3}{c}{Binding sfTA} \\ \hline
                System   & 1L      & 2L     & Bind.  & 1L        & 2L        & Bind.    & 1L        & 2L        & Bind.    \\ \hline
                Fig. \ref{fig:set1_original_sfta}--\ref{fig:set1_binding_plus_all} & 26(2)   & 20(2)  & 20(3)  & 21(2)     & 20(2)     & 6(1)     & 25(2)     & 23(2)     & 4(1)     \\
                \reffig{fig:set2_all}   &--       & --      & --      & 18(4)     & 23(4)     & 16(1)    & 14(4)     & 21(4)     & 10(1)    \\ \hline
            \end{tabular}
        \end{table}

        Our goal in this section is to analyze three sfTA variants to determine their accuracy for obtaining the interlayer binding correlation energy of bilayer systems. The three variants are defined as follows:
        \begin{itemize}
            \item \textbf{sfTA:} The original method applied to the 1L and 2L systems separately.There are different (unpaired) twist angles between the 1L and 2L systems, and both systems use the transition structure factor to find the special twist angle.
            \item \textbf{paired sfTA:} The 1L and 2L systems now use the same set of random twist angles but use the transition structure factor in the same way as original sfTA.
            \item \textbf{binding sfTA:} The 1L and 2L systems use the same set of random twist angles from paired sfTA, but now the special twist for both systems is determined using the binding transition structure factor.
        \end{itemize}
        These variants are applied to a small test set of five test systems, all of which have hexagonal ($P6/mmc$) symmetry: \ce{C}, \ce{GaN}, \ce{BN}, \ce{SiC}, and \ce{Si}.
        For the original sfTA variant, we anticipate there may be reduced accuracy due to the change in dimensionality, as well as the difference is supercell sizes used in both works.
        We expect that the modifications made to the paired and binding sfTA variants will have little to no impact on the accuracy of the results compared to the original sfTA method for the 1L and 2L systems, but that the binding energy will show improvement for both paired and binding sfTA.

        \reffig{fig:set1_original_sfta} shows the correlation energies for the original sfTA method applied to the mono- (1L) and bilayer (2L) systems for the small test set, as well as the interlayer binding correlation energy as described by \refeq{eq:binding_energy}.
        All energies are plotted as a difference to the $\Gamma$-point energy, marked at $0$ (dashed black line).
        For CCSD, individual twist angle energies are shown as the pink box-and-whisker plots, TA energies with standard error are shown as dark blue boxes, and sfTA energies are shown as dark green lines. 
        
        From the figure, we can see that the sfTA energies are not in agreement with the TA energy within one standard error for any of the systems.
        Overall, across all systems, the mean absolute errors (MAE) are 26(2) meV/atom for 1L, 20(2) meV/atom for 2L, and 20(3) meV for the interlayer binding energy, as shown in \reftab{tab:MAE}.
        These MAE are all larger than the value of 3.9(8) meV/atom from the original sfTA paper.\cite{mihm_shortcut_2021}
        As previously mentioned, this is not unexpected because the materials in that paper were bulk 3D materials, while our materials are 2D.\cite{mihm_shortcut_2021}
        In addition, we are using only $1\times1\times1$ unit cells, while they used $2\times2\times2$ supercells, which can lead to a difference of finite size errors present in the system.\cite{mihm_shortcut_2021}
        Overall, these results are less accurate than the original sfTA paper, but this is acceptable as these are only being used a baseline to compare against our two new variants.\cite{mihm_shortcut_2021}

        \reffig{fig:set1_paired_sfta} shows the correlation energies for the paired sfTA method.
        Across all systems, the MAE from \reftab{tab:MAE} are 21(2) meV/atom for 1L, 20(2) meV/atom for 2L, and 6(1) meV for the interlayer binding energy.
        The MAE for paired sfTA for the 1L and 2L systems are very similar to those from original sfTA in \reffig{fig:set1_original_sfta}, while the MAE for the interlayer binding energy shows a decrease of 14(3) meV/atom.\cite{mihm_accelerating_2021}
        These observations make it clear that paired sfTA leads to improved comparison to TA in the interlayer binding energy, while the 1L and 2L energies are largely unaffected.

        \reffig{fig:set1_binding_plus_all} shows the interlayer binding correlation energies for original, paired, and binding sfTA for the small test set.
        All sfTA variants are shown for ease of comparison.
        Binding (orange dashed line) and paired (dark green line) sfTA data are shown to the left and labeled as "paired" methods, while original sfTA (dark green line) is to the right, labeled as "unpaired".
        We can see from the figure that the binding sfTA energy is in agreement with the TA energy (dark blue box) within one standard error for both \ce{BN} and \ce{Si}.
        The binding sfTA results are also consistently closer to TA than paired sfTA in the figure.
        The MAE for binding sfTA across all systems as denoted in \reftab{tab:MAE} is 25(2) meV/atom for 1L, 23(2) meV/atom for 2L, and 4(1) meV/atom for the binding energy.
        These values all agree with the paired sfTA results within the calculated standard error, though the binding MAE is once again reduced, this time by 2 meV/atom.
        Notably, the MAE for the binding energy is now in agreement with the result from the original sfTA paper of 3.9(8) meV/atom.\cite{mihm_shortcut_2021}
        Overall, binding sfTA shows a minor improvement over paired sfTA, but this small difference leads to the same agreement between sfTA and TA for bilayer materials as was shown in the literature for bulk materials.
        
    \subsection{Binding sfTA for Challenging Systems}
        \label{sec:results:challenge}
        \begin{figure*}
            \centering
            \includegraphics[width=\linewidth]{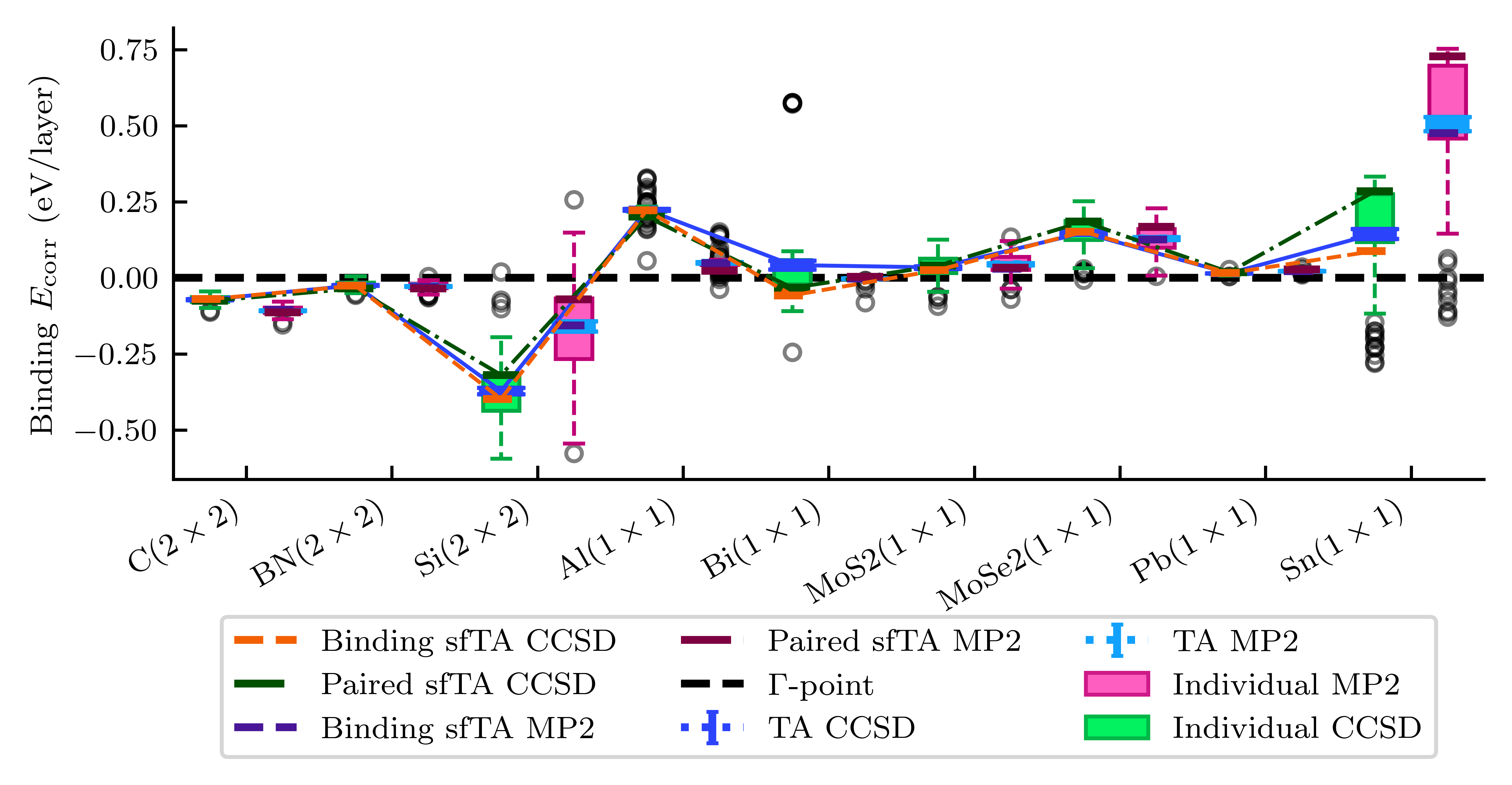}
            \caption{Binding correlation energy using TA, paired sfTA, and binding sfTA for challenge test set. Systems use either $1\times1$ or $2\times2$ supercells as noted. Results are shown as a difference to $\Gamma$-point. Lines connecting CCSD energy values across systems are visual aids for quick reference. Calculations were on a monolayer and bilayer, then binding energies were calculated using Eq. \ref{eq:binding_energy}. Mean absolute error (MAE) values relative to TA are 39 meV/layer (paired sfTA) and 23 meV/layer (binding sfTA).}
            \label{fig:set2_all}
        \end{figure*}

        \reffig{fig:set2_all} shows the binding correlation energies from the paired sfTA (dark green) and binding sfTA (orange) variants and the TA result (dark blue) for nine additional systems: \ce{C} ($2\times2$), \ce{BN} ($2\times2$), \ce{Si} ($2\times2$), \ce{Al} ($1\times1$), \ce{Bi} ($1\times1$), \ce{MoS2} ($1\times1$), \ce{MoSe2} ($1\times1$), \ce{Pb} ($1\times1$), and \ce{Sn} ($1\times1$).
        CCSD TA and sfTA variant energies are connected by lines between systems to help draw comparison.

        We can see in the figure that paired sfTA, binding sfTA, and the TA result are close together for most systems.
        In some cases, like \ce{Si} ($2\times2$) and \ce{Sn} ($1\times1$), there is some separation between the variants.
        In those cases, though, it is clear that the binding sfTA results is consistently closer to the TA result.
        This is supported by \reftab{tab:MAE} which shows MAEs of 16(1) meV/atom and 10(1) meV/atom for paired and binding sfTA, respectively.
        While these values are larger than what seen for the smaller test system, it is still within what is generally considered good agreement for highly accurate correlation methods.
        Therefore, the sfTA variants for the interlayer binding energy show good agreement with TA for a wide variety of systems.

    \subsection{Binding Energy Landscapes}
        \label{sec:results:contour}
        \begin{figure}
            \centering
            \includegraphics[width=0.95\linewidth]{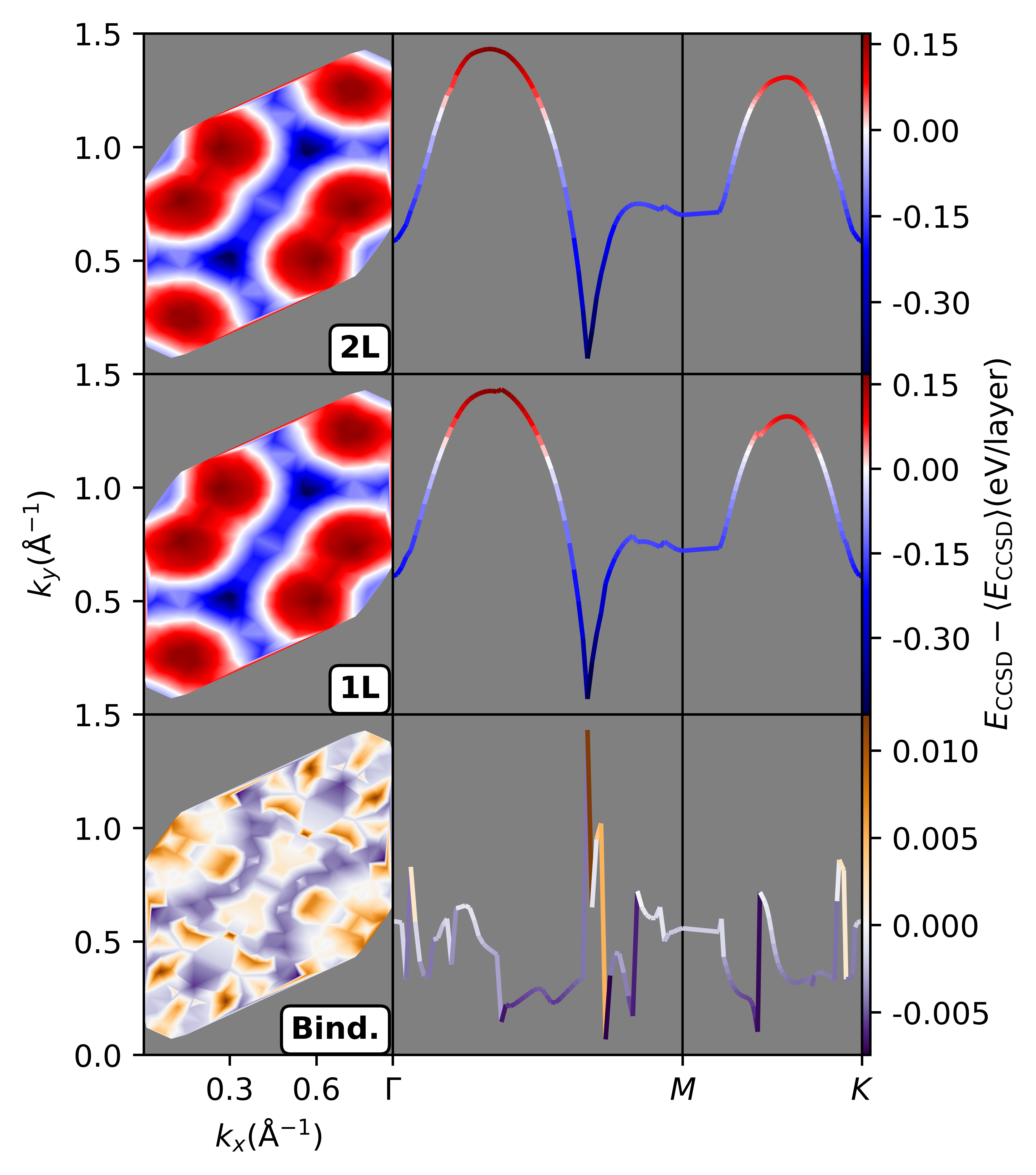}
            \caption{
                Contour plots of the monolayer (1L), bilayer (2L) and binding (Bind.) CCSD correlation energy for graphene as a function of the two-dimensional twist angle offset $\left(k_x,k_y\right)$.
                Energies are differenced to the TA result (white) and are reported per-layer.
                Also shown are slices along the high-symmetry path $\Gamma\rightarrow M\rightarrow K$, corresponding to the fractional coordinates $\left(0,0\right)$, $\left(\sfrac{1}{2},\sfrac{1}{2}\right)$, and $\left(1,0\right)$, respectively.
                The slices were plotted using $2^7+1\ (129)$ additional calculations exactly along the path.
                These figures show clear, symmetric features for the isolated 1L and 2L systems, with very slim boundaries of energy close to TA.
                They also show the binding energies are much flatter, being smaller in magnitude and generally closer to TA.
                }
            \label{fig:contour_carbon}
        \end{figure}

        While the energy distributions shown in previous figures help demonstrate that the sfTA variants are successful in locating near-TA twist angles, twists that are representative of the TA energy, it remains unclear how and why they are selecting different twist angles from each other.
        To investigate this, Fig. \ref{fig:contour_carbon} shows contour plots of the mono- (1L) and bilayer (2L) energies, as well as the interlayer binding energy (Bind.), as a function of the two-dimensional twist angle $\left(k_x,k_y\right)$ for the $1\times1$ graphene system previously shown in Figs. \ref{fig:set1_original_sfta}--\ref{fig:set1_binding_plus_all}. %
        Twists used to construct the contours are the same as before, though here they have been duplicated according to reciprocal lattice symmetry for a fourfold increase in resolution.
        Energies are shown as a difference to the TA energy.
        For the 1L and 2L systems, low energies are blue, high energies are red, and the TA energy is white.
        Binding energies use a different color scale (brown to violet) to emphasize the different energy type and scale.
        Next to the contour plots are diagonal slices along the $\Gamma\rightarrow M\rightarrow K$ high-symmetry path, which were constructed from $2^7+1\ (129)$ additional calculations, for detail.

        The 1L and 2L contour plots appear nearly identical.
        Both have clear regions of high-energy twist angles (red) separated from low-energy twist angles (blue) with thin boundaries of near-TA energy twists (white) separating them.
        As can be seen from the high-symmetry paths, the TA energies are consistently on a region of the contour surface with a steep slope, meaning the corresponding twists are easy to miss, which supports the need for improved twist-angle sampling techniques.
        We also see that the high symmetry points $\Gamma$, $M$, and $K$ are consistently below TA, on the order of \textasciitilde150 meV/atom, making them poor choices for the special twist angle.
        The behavior at the high-symmetry twists is somewhat expected, as it is well-known that $\Gamma$-point can be very inaccurate and lead to convergence issues for bulk materials, though as we will see, this phenomenon is muted for the binding correlation energy.

        \begin{table}[]
            \caption{
                Maximum absolute error relative to the TA binding correlation energy using consistent twist angles between 1L and 2L for all systems, in meV/atom. 
                All systems are $1\times1$ unless noted otherwise.
                }
            \label{tab:max_error}
            \begin{tabular}{cc|cc}
                \hline
                \multicolumn{2}{c|}{Small Set} & \multicolumn{2}{c}{Challenging Set} \\ \hline
                System       & Max Error     & System     & Max Error    \\ \hline
                \ce{C}            & 6.1(2)        & \ce{C} $\left(2\times2\right)$          & 20.7(6)      \\
                \ce{BN}           & 21.5(5)       & \ce{BN} $\left(2\times2\right)$         & 16.4(6)      \\
                \ce{Si}           & 143(7)        & \ce{Si} $\left(2\times2\right)$         & 196(6)       \\
                \ce{SiC}          & 78(2)         & \ce{Al}         & 84(2)        \\
                \ce{GaN}          & 72(2)         & \ce{Bi}         & 267(7)       \\
                             &               & \ce{MoS2}       & 42(1)        \\
                             &               & \ce{MoSe2}      & 52(2)        \\
                             &               & \ce{Pb}         & 6.1(2)       \\
                             &               & \ce{Sn}         & 212(8)       \\ \hline
                Average      & 64(8)         & Average      & 100(10)      \\ \hline
            \end{tabular}
        \end{table}
        
        The binding energy contour plot appears very different from the 1L and 2L plots, with the hexagonal symmetry no longer clearly visible.
        The contour plot is much flatter than the 1L and 2L plots above, with almost the entire region being near-TA (white).
        The remaining high- (brown) and low-energy (violet) regions are also much lower than their corresponding energies for 1L and 2L.
        This suggests that across all twist angles, there is a cancellation of errors, meaning a poor choice of twist angle for the 1L or 2L system could still give near-TA binding energies.
        This is further supported by looking at the diagonal slice.
        For the binding energy, the previously inaccurate high-symmetry $\Gamma$, $M$, and $K$ points are now very close to TA.
        Even the most extreme value for the binding energy, corresponding to a low-energy valley for the 1L and 2L systems, is off from TA by only 6.1(2) meV/atom for this system, as shown in \reftab{tab:max_error}, meaning \textit{any} choice of twist angle should result in near-TA binding energies so long as the twist is constant between 1L and 2L.
        This is not true for all systems, though, especially those with large outliers, like \ce{Bi}, or those whose energies span a large range, like \ce{Si}, both of which are visible in \reffig{fig:set2_all}.

\section{Conclusions}
    In this work, we showed that the paired sfTA method resulted in closer agreement to TA compared to original sfTA for the interlayer binding energies of bilayer materials, including metals, semiconductors, and insulators.
    We also showed that the binding sfTA variant further improves upon paired sfTA.
    For the small test set (Figs. \ref{fig:set1_original_sfta}--\ref{fig:set1_binding_plus_all}), MAE values for the interlayer binding energy for each variant were 20(3) meV/atom (original sfTA), 6(1) meV/atom (paired sfTA), and 4(1) meV/atom (binding sfTA), supporting the conclusion that binding sfTA leads to the closest agreement with TA among the variants.
    This was further supported by a second, larger test set (\reffig{fig:set2_all}, with MAE values of 16(1) meV/atom (paired sfTA) and 10(1) meV/atom (binding sfTA).
    Although the MAE results from the more challenging test set were larger, they are of a comparable magnitude to the previous results and show a similar level of improvement from paired to binding sfTA.
    Taken together, both test sets show that our sfTA variants improve the accuracy of the binding correlation energy while retaining the cost-saving benefits of original sfTA.

    We further explored the dependence of the correlation energy, including the binding correlation energy, on the two-dimensional twist angle.
    Looking at contour plots of $1\times1$ graphene we showed that when the twist angle is held constant between the 1L and 2L systems used to calculate the binding energy, like what is done for binding sfTA, the resulting binding correlation energies are consistently near-TA.
    This behavior appears to be from a cancellation of errors, and holds even for twists that were far from TA for the 1L or 2L system, such as $\Gamma$-point.
    
    While the results here have improved our ability to perform CCSD on 2D systems, there are still some lingering questions.
    First, while the transition structure factor has been critical for this work, there is the question of if there are alternate electronic structure properties that might better suited for twist angle selection for low-dimensional systems.
    For example, before structure factor twist averaging, we explored connectivity twist averaging, where we select twist angles based on eigenvalue data, as well as energy matching. \cite{mihm_optimized_2019,mihm_accelerating_2021}
    It may be worthwhile to explore these options again for low-dimensional materials.
    Additionally, there is a question of if the variants used here can be easily applied to different types of low-dimensional systems, like small molecules adsorbed to metal surfaces.

    Overall, this project has improved our understanding of how the transition structure factor can be used for interacting systems, as well as anisotropic systems with periodic boundary conditions.
    This work has expanded the applicability of the original sfTA method to more systems.\cite{mihm_accelerating_2021} 
    We expect this work to be transferrable for many other anisotropic systems with binding interactions.

\section*{Supplementary Materials}
    Lattice parameters and interlayer distances for structures used in this work are provided in the supplementary information.

\begin{acknowledgments}
Research was primarily supported by the National Science Foundation under Award Number CHE-2045046. The authors thank Michigan State University for their support. We also thank MSU and the University of Iowa for computational resources used in this project. We thank Tina Mihm for comments on the manuscript. Special thanks to Andreas Gr{\"u}neis and Tobias Sch{\"a}fer from TU Wien, Vienna, Austria, for optimized structures for \ce{C} and \ce{BN}, and for helpful discussions.
\end{acknowledgments}

\section*{Author Declarations}
    \subsection*{Conflict of Interest}
        The authors have no conflicts to disclose
    \subsection*{Author Contributions}
        \textbf{Ryan A. Baker}: investigation (lead); conceptualization (equal); formal analysis (lead); visualization (lead); writing -- original draft (lead); writing -- review and editing (equal). 
        \textbf{William Z. Van Benschoten}: conceptualization (equal); writing -- review and editing (equal). 
        \textbf{James J. Shepherd}: conceptualization (lead); formal analysis (equal); supervision (lead); writing -- original draft (equal); writing -- review and editing (equal).

\section*{Data Availability}
    The data that support the findings of this article are available in the article and its supplementary material.

 \end{document}